\documentclass[aps,floatfix]{revtex4}

\usepackage{hyperref,hypcap,ae,tikz,color}
\usepackage{latexsym}
\usepackage{graphicx}
\usepackage{xcolor}

\usepackage[T1]{fontenc}
\usepackage[latin9]{inputenc}

\usepackage{mathrsfs}
\usepackage{amsmath}
\usepackage{amssymb}

\hypersetup{
	colorlinks=true,
	citecolor=green,
    linkcolor=magenta,
    urlcolor=blue
	}

\newcommand{\be}{\begin{equation}}
\newcommand{\ee}{\end{equation}}
\newcommand{\bea}{\begin{eqnarray}}
\newcommand{\eea}{\end{eqnarray}}

\begin{document}
\title{Structure Scalars for Charged Dissipative Spherical Collapse in $f(R,T)$ Gravity}
\author{Uttaran Ghosh and Sarbari Guha}
\affiliation{Department of Physics, St. Xavier's College (Autonomous), Kolkata 700016, India}

\begin{abstract}
We examine the structure scalars constructed from the orthogonal splitting of the Riemann tensor for the spacetime metric describing the interior of a charged matter configuration undergoing dissipative collapse in the framework of $f(R,T)$ gravity (where $R$ and $T$ are the Ricci scalar and
the trace of energy-momentum tensor, respectively), and also the way these quantities influence the various physical parameters of the collapsing matter. In absence of dissipation, the energy density inhomogeneity is found to be influenced by the structure scalar $X_{TF}$ and the mass-function of the collapsing matter. Further, the presence of charge affects the structure scalars and the total mass-energy content. The dependence of the various physical parameters like heat dissipation, energy density inhomogeneity, evolution of the expansion scalar, the shear scalar, effective homogeneous energy density, and pressure anisotropy on the structure scalars, have been clearly indicated along with a discussion on the complexity factor of the collapsing configuration. The $f(R,T)$ junction conditions have been presented, showing the matching conditions for the matter Lagrangian and their derivatives at the boundary. The energy conditions are also presented and the possibility of violation of the Strong Energy Condition has been discussed.
\end{abstract}

\maketitle

\section{Introduction}
Investigations on the determination of the final fate of a stellar object collapsing under its own gravity gained prominence among researchers ever since the seminal work on the collapse of a spherical dust ball by Oppenheimer and Snyder \cite{OS}, and independently by Datt \cite{Datt} during 1938-39. The general relativistic consideration of a matter combination represented by a rank two energy-momentum tensor left open a number of possibilities for the end result of gravitational collapse, depending on the type of collapsing matter. Details of the involvement or absence of shear viscosity, pressure anisotropy, electric field, dissipative effects like heat flow, play a huge role in deciding whether the final fate of collapse will be a black hole or a naked singularity, a regime in which the usual laws of physics cannot be applied any more. Over the years, researchers have studied a number of such cases in detail in the context of General Relativity (GR) \cite{Glass1981, Santos1985, Herrera_et_al2009, OlivSan1985, OliPachSantos1986, HerDenSan1989, ChanKicheDenSan1989, Chan1993, ChanHerSan1994, Chan2000, HerSanPRD04, HerSanDenGRG2012,OlivSan1987}.

In 2009, Herrera et al \cite{Herrera_et_al2009} demonstrated for the first time that a set of scalars derived from the orthogonal splitting of the Riemann tensor have distinct physical interpretation and are particularly suited for the characterisation of self-gravitating relativistic fluids. They termed these scalars as ``structure scalars'', which influence the various physical parameters of the collapsing matter. Their construction followed the consideration of Bel \cite{Bel1961}, who introduced for the first time the idea of the orthogonal splitting of the Riemann tensor by defining the dual tensors corresponding to the Riemann tensor, and led to the formulation of the gravitational super energy tensor. In 2008, G\'{o}mez-Lobo \cite{Gomez_Lobo_CQG2008} in an effort to shed light on the true physical meaning of superenergy, performed the orthogonal splitting of the Bel and Bel-Robinson tensors, and analysed the different parts arising out of the splitting. A detailed presentation of the orthogonal splitting of the Riemann tensor was also provided in this paper.

Following the work of Herrera et al \cite{Herrera_et_al2009}, several authors have characterized the evolution of self-gravitating systems in terms of structure scalars, in the context of General Relativity and also in other modified theories of gravity. Herrera, Di Prisco and Ibanez investigated the role of electric charge and the cosmological constant in the structure scalars \cite{HerreraPRD2011}. Herrera, Di Prisco and Ospino studied structure scalars for relativistic fluids with cylindrical symmetry \cite{HerreraGRG2012}. Sharif and Bhatti studied structure scalars for charged relativistic fluids in cylindrical symmetry \cite{Sharif2012}. They also studied the role of structure scalars in presence of charge in plane symmetry \cite{SharifMPLA2012}, charged static solutions with axial symmetry \cite{SharifAPSS2014}, structure scalars for tilted Szekeres geometry and also the super-Poynting vector \cite{SharifIJMPD2015}.
Sharif and Manzoor analysed the role of structure scalars and obtained the inhomogeneity factors for the spherical self-gravitating fluid models in Brans-Dicke theory, and investigated the spherical static anisotropic solutions with inhomogeneity using these scalars \cite{SharifPRD2015}. They also studied the role of structure scalars in cylindrical systems in Brans-Dicke gravity, and showed that cylindrical systems must necessarily be inhomogeneous \cite{SharifManzoorAPSS2015}.

The fact that the universe is undergoing a late-time accelerated expansion emerged from the analysis of the data obtained from the observation of the Type Ia spernova \cite{SSTC, Perlmutter}. Explanation for this accelerated expansion is provided by assuming the presence of a mysterious component called ``dark energy'', about which we do not have much information, except that it constitutes a large part of the total matter-energy content of the universe. Further, the galactic rotation curves show the presence of ``dark matter'' which forms the major constituent of the matter component of the universe. A modification of the Einstein-Hilbert action in GR, by replacing the Ricci scalar $R$ with a function of the Ricci scalar $f(R)$, yields us the $f(R)$ theory of gravity. This modification gives rise to some extra terms related to the curvature of the spacetime, which are purely geometrical in origin. These terms can be considered to explain the accelerated expansion of the universe, without the need to invoke the cosmological constant $\Lambda$, which is sometimes used in the Einstein Field Equations in GR for the same purpose. The works by De Felice and Tsujikawa \cite{DeFelice2010}, Sotiriou and Faraoni \cite{Sotiriou2010}, and Capozziello and Laurentis \cite{Capozziello2011} provide further insights into modified gravity theories. Modified gravity theories allowing a unification of early time inflation and late-time acceleration was also discussed in \cite{NojiriOdintsov2011}. Development of modified gravity theories in the context of inflation, bounce and late time acceleration was discussed in \cite{NojiriOdintsov2017}. A comparative study of gravitational collapse in General Relativity and $f(R)$ gravity, or more precisely, $R^{2}$ gravity can be found in \cite{AstashenokIJGMMP2019}. Sharif and Yousaf studied structure scalars in the case of radiating cylindrical collapse in $f(R)$ gravity \cite{SharifAPSS2015}. Bhatti, Yousaf and Tariq studied structure scalars and their evolution for massive bodies in $f(R)$ gravity \cite{BhattiEPJC2021}, and analysed structure scalars in $f(R)$ gravity also in the presence of electric charge \cite{BhattiPhysScr2021}. Further, they examined the role of structure scalars in the evolution of compact objects in Palatini $f(R)$ gravity \cite{BhattiCJP2021}.

An extension of the $f(R)$ theory is the $f(R,T)$ theory of gravity formulated by Harko et al \cite{Harko2011}. In this theory the Ricci scalar $R$ in the Enstein-Hilbert action is replaced by a function of both $R$ and $T$, where $T$ is the trace of the energy-momentum tensor. Thus the $f(R,T)$ function depends both on the curvature of the spacetime, as well as the matter part. The inclusion of the trace takes into account the presence of imperfect exotic fluids or possible quantum effects such as the conformal anomaly. The extra curvature terms arising from the geometry can explain the accelerated expansion of the universe, without the need to invoke dark energy. In addition, the galactic rotation curves can also be explained without invoking dark matter. A suitable choice of the $f(R,T)$ function is a function which is linear in both $R$ and $T$. This linear form $R+\lambda T$ leads to power-law type of scale factors. Sahoo and his collaborators showed that this model of $f(R)+\lambda T$ gravity can be used as an alternative to the cosmic acceleration \cite{Sahoo2018}. Guha and Ghosh studied dynamical conditions and causal transport phenomena for a dissipative spherical collapse in $f(R, T)$ gravity \cite{GuhaGhosh2021}. They also studied the formation of singularity and apparent horizon for dissipative spherical collapse in $f(R, T)$ gravity \cite{GhoshGuha2025}. Yousaf et al \cite{Yousaf2016a} explored the evolution of compact objects in $f(R, T)$ gravity with the help of structure scalars for the case of spherical systems coupled with heat- and radiation-emitting shearing viscous matter distributions. They explicitly demonstrated that even in modified gravity, the evolutionary phases of relativistic stellar systems could be analyzed with the help of these scalar functions. In a separate paper \cite{Yousaf2016b} these authors examined irregularity factors for a self-gravitating spherical star evolving in the presence of an imperfect fluid in $f(R, T)$ gravity, where they demonstrated that, as the complexity of the matter with the anisotropic stresses increases, the inhomogeneity factor corresponds more closely to one of the structure scalars. Hussain et. al. studied the role of structure scalars in the $f(R,T)$ theory of gravity \cite{Hussain2017}. Yousaf, Bhatti and Farwa studied axial and reflection-symmetric self-gravitating systems and structure scalars in $f(R,T)$ gravity \cite{Yousaf2021}.

Any physical property of a collapsing system, apart from isotropic pressure and homogeneous energy density, adds to the complexity of the system. More precisely, parameters like energy density inhomogeneity, pressure anisotropy, dissipative and shearing effects, electromagnetic field, all add to the complexity of a collapsing fluid configuration, while the simplest system of a fluid with homogeneous energy density and isotropic pressure is said to possess a vanishing complexity. This new definition of complexity factor was introduced by Herrera \cite{Herrera2018} for static spherical self-gravitating systems, based on a quantity that appears in the orthogonal splitting of the Riemann tensor, in the context of general relativity. This proposal was immediately followed up by Herrera et al \cite{Herrera&ors2018}, who extended it to fully dynamic situation, where they also considered the condition of minimal complexity of the pattern of evolution. This treatment was applied further to other cases in the context of GR \cite{Herrera&ors2019a, Herrera&ors2019b}. Sharif and collaborators \cite{SharifButt2018, SharifTariq2020} also examined the complexity factor for dynamical spherical systems under various conditions. Complexity factor have also been discussed in $f(R,T,R_{\mu\nu}T^{\mu\nu})$ gravity \cite{YousafPDU2020}.

In this paper, we have chosen to work in the framework of $f(R, T)$ gravity. The arbitrariness in the choice of the $f(R,T)$ function leads to flexibility in the choice of models. It may lead to insights regarding the behaviour of gravity in high-energy scenarios by the introduction of quantum corrections in some of its functional forms. It may be used to model compact objects like black holes and neutron stars, and investigate their various properties including stability. It can also offer explanations to the formation of large-scale structures in the universe. The paper is divided into the following sections : In section II, a brief formalism of the $f(R,T)$ theory of gravity has been provided, followed by the definition of the structure scalars in section III. In section IV, the interior spacetime and the enrgy momentum tensor have been specified, and the structure scalars, the Weyl scalar, the expansion scalar and the shear scalar have been obtained in terms of the metric coefficients. The field equations have been presented in section V, and the relation between the physical matter variables and the structure scalars have been established in section VI, with subsection A discussing the constant $R$ and $T$ case for a dust ball, and subsection B showing the form of these relations for a linear $f(R,T)$ function, followed by a discussion on the complexity factor in subsection C. The exterior spacetime and the corresponding field equations have been presented in section VII, and the interior spacetime and the energy-momentum tensor in terms of the Ricci tensor in section VIII. The junction conditions in $f(R,T)$ gravity have been discussed in section IX. The energy conditions in $f(R,T)$ gravity have been discussed in section X. Finally, the results and conclusions are presented in section XI.
 	
\section{The $f(R,T)$ Formalism}

The modified Einstein-Hilbert action in $f(R,T)$ gravity, is given by
\begin{equation}\label{EHaction}
S=\int ~d^{4}x \sqrt{-g} \Bigg(\frac{f(R, T)}{16\pi G} + \mathcal{L} _ {m} \Bigg).
\end{equation}
The $f(R,T)$ field equations are obtained by variation of this action, and are given by
\begin{align}\label{frtfe}
G_{\mu\nu} &=\frac{1}{f_{R}}\left[\left(1+f_{T}\right)T_{\mu\nu}^{(m)}-L_{m}g_{\mu\nu}f_{T}\right. \nonumber \\
& \left.+\frac{1}{2}\left(f-Rf_{R}\right)g_{\mu\nu}-D_{\mu\nu}\right],
\end{align}

where $g_{\mu\nu}$ is the metric tensor representing the four-dimensional spacetime in the region interior to the collapsing matter, $G_{\mu\nu}$ is the Einstein tensor, $R$ is the Ricci scalar, $T$ is the trace of the energy-momentum
tensor, $f_{R}$ and $f_{T}$ are the derivatives of the $f(R,T)$
function with respect to $R$ and $T$, respectively, $L_{m}$ is
the interior matter Lagrangian, and $D_{\mu\nu}=\left(g_{\mu\nu}\Box-\nabla_{\mu}\nabla_{\nu}\right)f_{R}$
which includes the higher order curvature terms, which acts as the
source of dark energy.
Before describing the interior spacetime for our investigation of the collapse, we formulate the structure scalars, which play an important role in influencing the physical parameters of the collapsing matter.

\section{Structure Scalars}

Following Bel \cite{Bel1961} and Herrera \cite{Herrera_et_al2009}, we formulate the quantities termed ``structure scalars" by an orthogonal splitting of the Riemann tensor $R_{\alpha\beta\gamma\delta}$ for the interior spacetime. A prescription for this orthogonal splitting was given by Gomez-Lobo \cite{Gomez_Lobo_CQG2008}. In \cite{Herrera_et_al2009}, Herrera showed that in the framework of GR, these scalar quantities can be utilised to completely describe the structure and evolution of the spherically symmetric anisotropic dissipative fluid which is self-gravitating. These scalars are found to influence the physical properties of the fluid, such as, the energy density, pressure anisotropy, heat flux, and the active gravitational mass, and combinations of these scalars are sufficient to express the solutions to the Einstein Field Equations in the static case.

In analogy with Herrera \cite{Herrera_et_al2009}, the right dual, left dual, and double dual of the Riemann
tensor are specified as follows :
\begin{equation}\label{rdual}
R_{\alpha\beta\gamma\delta}^{*}=\frac{1}{2}\eta_{\epsilon\rho\gamma\delta}R_{\ \ \alpha\beta}^{\epsilon\rho}
\end{equation}

\begin{equation}\label{ldual}
^{*}R_{\alpha\beta\gamma\delta}=\frac{1}{2}\eta_{\alpha\beta\epsilon\rho}R_{\ \ \gamma\delta}^{\epsilon\rho}
\end{equation}

\begin{equation}\label{ddual}
^{*}R_{\alpha\beta\gamma\delta}^{*}=\frac{1}{2}\eta_{\alpha\beta}^{\ \ \epsilon\rho}R_{\epsilon\rho\gamma\delta}^{*}
\end{equation}
where, $\eta_{\alpha\beta\gamma\delta}$ is the four-index Kronecker delta symbol.

The spacetime metric is described by the tensor $g_{\mu\nu}$ and the four-velocity vector is given by
$u^{\mu}$. Utilising equations \eqref{rdual}, \eqref{ldual} and \eqref{ddual}, and following the prescription by Herrera \cite{Herrera_et_al2009}, we have :
\begin{equation}\label{yab}
Y_{\alpha\beta}=R_{\alpha\gamma\beta\delta}u^{\gamma}u^{\delta}
\end{equation}

\begin{equation}\label{xab}
X_{\alpha\beta}=^{*}R_{\alpha\gamma\beta\delta}^{*}u^{\gamma}u^{\delta}=\frac{1}{2}\eta_{\ \ \alpha\gamma}^{\epsilon\rho}R_{\epsilon\rho\beta\delta}^{*}u^{\gamma}u^{\delta}
\end{equation}

\begin{equation}\label{zab}
Z_{\alpha\beta}=^{*}R_{\alpha\gamma\beta\delta}u^{\gamma}u^{\delta}=\frac{1}{2}\eta_{\alpha\gamma\epsilon\rho}R_{\ \ \beta\delta}^{\epsilon\rho}u^{\gamma}u^{\delta}
\end{equation}

Denoting the trace part and trace-free
part of $X_{\alpha\beta}$ by $X_{T}$ and $X_{TF}$ respectively, and the trace part and trace-free part of $Y_{\alpha\beta}$ by $Y_{T}$ and $Y_{TF}$
respectively,
with $\chi_{\mu}$ as a unit vector in the radial direction, we can
write, using the projection tensor $h_{\alpha\beta}=g_{\alpha\beta}+u_{\alpha}u_{\beta}$,
\begin{equation}\label{xab1}
X_{\alpha\beta}=\frac{1}{3}X_{T}h_{\alpha\beta}+X_{TF}\left(\chi_{\alpha}\chi_{\beta}-\frac{1}{3}h_{\alpha\beta}\right)
\end{equation}

and,
\begin{equation}\label{yab1}
Y_{\alpha\beta}=\frac{1}{3}Y_{T}h_{\alpha\beta}+Y_{TF}\left(\chi_{\alpha}\chi_{\beta}-\frac{1}{3}h_{\alpha\beta}\right)
\end{equation}
The trace parts and trace-free parts of these vectors are referred to as structure scalars since it can be shown from the field equations for the interior spacetime that the structure scalars influence various matter-energy parameters like the inhomogeneity in the matter-energy density, shear viscosity, pressure anisotropy, electric field and so on.

\section{The Interior Spacetime and the Energy-Momentum Tensor}

The metric describing the interior spacetime is the most general spherically
symmetric metric which is
\begin{align}\label{intmetric}
ds_{-}^{2} &=-A(r,t)^{2}dt^{2}+B(r,t)^{2}dr^{2} \nonumber \\
& +C(r,t)^{2}\left(d\theta^{2}+\sin^{2}\theta d\phi^{2}\right)
\end{align}

The matter in the interior spacetime is considered to be an anisotropic fluid undergoing dissipation in the form of heat flux and free-streaming radiation, with shear viscosity, whose energy-momentum tensor is given by

\begin{align}\label{intEM}
T_{\mu\nu} &=\left(\rho+p_{\perp}\right)u_{\mu}u_{\nu}+p_{\perp}g_{\mu\nu}+\left(p_{r}-p_{\perp}\right)\chi_{\mu}\chi_{\nu}\nonumber \\
& +q_{(\mu}u_{\nu)}+\epsilon l_{\mu}l_{\nu}-2\eta\sigma_{\mu\nu}
\end{align}
where, $\rho$ is the interior energy density, $p_{\perp}$ is the tangential pressure, $p_{r}$ is the radial pressure, $u^{\mu}$ is the four-velocity, $\chi^{\mu}$ is a unit vector in the radial direction, $q_{\mu}$ is the heat flux vector in the radial direction and is given by $q_{\mu}=q\chi_{\mu}$, with $q$ being the heat flux, $\epsilon$ is the free-streaming radiation density, $l^{\mu}$ is a null 4-vector, $\eta$  is the coefficient of shear viscosity, which is positive, and $\sigma_{\mu\nu}$ is the shear tensor. The following conditions are satisfied by these tensors :
\begin{align}
u^{\mu}u_{\mu}=-1, \;\; u^{\mu}q_{\mu}=0, \;\; \chi^{\mu}\chi_{\mu}=1, \;\; \nonumber \\
\chi^{\mu}u_{\mu}=0, \;\; l^{\mu}u_{\mu}=-1, \;\; l^{\mu}l_{\mu}=0.
\end{align}

Assuming the observer to be comoving with the collapsing matter, we have
\begin{align}
u^{\mu}=A^{-1}\delta^{\mu}_{0}, \;\; \chi^{\mu}=B^{-1}\delta^{\mu}_{1},  \;\; \nonumber \\
l^{\mu}=A^{-1}\delta^{\mu}_{0}+B^{-1}\delta^{\mu}_{1}.
\end{align}
It is further considered that the collapsing matter ball is charged, as a result of which, Maxwell's electromagnetic stress-energy tensor needs to be taken in consideration, in addition to the usual matter energy-momentum tensor.
The electromagnetic stress-energy tensor is given by
\begin{equation}
E_{\mu\nu}=\frac{1}{4\pi}\left(-F_{\mu}^{\gamma}F_{\gamma\nu}+\frac{1}{4}F^{\delta\gamma}F_{\delta\gamma}g_{\mu\nu}\right)
\end{equation}

where, $F_{\mu\nu}=\varphi_{\nu,\mu}-\varphi_{\mu,\nu}$ is the Maxwell's
electromagnetic field tensor, with $\varphi_{\mu}$ being the electromagnetic
four-potential, which, in the absence of magnetic field has the scalar
potential $\varphi$ as its only non-zero component. The components
representing the vector potential all vanish in the absence of magnetic
field.

The effective energy-momentum tensor
is given by
\begin{align}
T_{\mu\nu}^{eff} &=\frac{1}{f_{R}}\left[\left(1+f_{T}\right)T_{\mu\nu}-L_{m_{int}} g_{\mu\nu}f_{T}\right. \nonumber \\
& \left.+\frac{\left(f-Rf_{R}\right)}{2}g_{\mu\nu}-D_{\mu\nu}^{-}+8\pi E_{\mu\nu}\right]
\end{align}

where $D_{\mu\nu}=\left(g_{\mu\nu}\Box-\nabla_{\mu}\nabla_{\nu}\right)f_{R}$
represents the dark energy source terms, and the negative sign in the superscript denotes the interior spacetime.
The Maxwell field equations are given by
\begin{equation}\label{maxwell1}
F_{[\mu\nu;\gamma]}=0 ,
\end{equation}
and,
\begin{equation}\label{maxwell2}
F^{\mu\nu}_{;\nu}=4\pi J^{\mu}.
\end{equation}
where $J^{\alpha}=ju^{\alpha}$ represents the four-current, with $j$ being the current density. The Maxwell field equations are
\begin{equation}\label{maxwellfe1}
\varphi''+\varphi'\left(-\frac{A'}{A}-\frac{B'}{B}+\frac{2C'}{C}\right)=4\pi AB^{2}j ,
\end{equation}
\begin{equation}\label{maxwellfe2}
\dot{\varphi}'-\varphi'\left(\frac{\dot{A}}{A}+\frac{\dot{B}}{B}-\frac{2\dot{C}}{C}\right)=0
\end{equation}

The conservation of charge requires
\begin{equation}\label{chargeconsvn}
J^{\mu}_{;\mu}=0,
\end{equation}

which yields the total charge, $Q(r)$ inside the collapsing sphere as
\begin{equation}\label{Q}
Q(r)=4\pi\int_{0}^{r}{jBC^{2}dr}.
\end{equation}
It can also be seen that
\begin{equation}\label{phiQrel}
 \frac{\varphi'^{2}}{A^{2}B^{2}}=\frac{Q^{2}}{C^{4}} .
\end{equation}

Following the works of other authors \cite{DiPrisco_PRD2007, Sharif_GRG2018, Ahmed_CJP2020}, the mass-function, which describes the total energy contained inside a 3D timelike hypersurface acting as the boundary of the collapsing matter, is given by
\begin{equation}\label{massfn}
m=\frac{C}{2}\left(1+\frac{\dot{C}^{2}}{A^{2}}-\frac{C'^{2}}{B^{2}}\right)+\frac{Q^{2}}{2C}
\end{equation}

The expansion scalar $\Theta$ is given by
\begin{equation}\label{expsn}
\Theta=\frac{1}{A}\left(\frac{\dot{B}}{B}+\frac{2\dot{C}}{C}\right),
\end{equation}

and the shear scalar $\sigma$ is defined as
\begin{equation}\label{shear}
\sigma=-\frac{1}{3A}\left(\frac{\dot{B}}{B}-\frac{\dot{C}}{C}\right).
\end{equation}

The structure scalars in terms of the metric coefficients, the function $m$ and the shear and expansion scalars, are given by
\begin{equation}\label{xt}
X_{T}=\frac{2\dot{C}\dot{B}}{CBA^{2}}-\frac{2C''}{CB^{2}}+\frac{2C'B'}{CB^{3}}+\frac{2m}{C^{3}}-\frac{Q^{2}}{C^{4}}
\end{equation}

\begin{equation}\label{xtf}
X_{TF}=-\frac{\dot{C}\dot{B}}{CBA^{2}}+\frac{C''}{CB^{2}}-\frac{C'B'}{CB^{3}}+\frac{2m}{C^{3}}-\frac{Q^{2}}{C^{4}}
\end{equation}

\begin{align}\label{yt}
Y_{T} &=\frac{1}{B^{2}}\left(\frac{A''}{A}-\frac{A'B'}{AB}+\frac{2A'C'}{AC}\right) \nonumber \\
& -\frac{1}{A^{2}}\left(\frac{\ddot{B}}{B}+\frac{2\ddot{C}}{C}\right)+\frac{\dot{A}\Theta}{A^{2}}
\end{align}

\begin{align}\label{ytf}
Y_{TF} &=\frac{1}{B^{2}}\left(\frac{A''}{A}-\frac{A'B'}{AB}-\frac{A'C'}{AC}\right) \nonumber \\
& -\frac{1}{A^{2}}\left(\frac{\ddot{B}}{B}-\frac{\ddot{C}}{C}\right)-\frac{3\dot{A}\sigma}{A^{2}}
\end{align}

The electric part of the Weyl tensor $C_{\alpha\beta\gamma\delta}$ is given by $\mathcal{E}_{\alpha\beta}=C_{\alpha\gamma\beta\delta}u^{\gamma}u^{\delta}$, and the magnetic part is given by $H_{\alpha\beta}=\frac{1}{2}\eta_{\alpha\gamma\epsilon\rho}C^{\epsilon\rho}_{\beta\delta}u^{\gamma}u^{\delta}$.
For the spherically symmetric case, all components of $H_{\alpha\beta}$ are zero and the magnetic part of the Weyl tensor vanishes.

The electric part of the Weyl tensor can also be expressed in terms of the Weyl scalar $\mathcal{E}$ as
\begin{equation}\label{Weylelectric}
\mathcal{E}_{\alpha\beta}=\mathcal{E}\left(\chi_{\alpha}\chi_{\beta}-\frac{1}{3}h_{\alpha\beta}\right)
\end{equation}

It satisfies the properties :
\begin{equation}
\mathcal{E}^{\alpha}_{\alpha}=0, \;\; \mathcal{E}_{\alpha\beta}=\mathcal{E}_{\left(\alpha\beta\right)}, \;\;\mathcal{E}_{\alpha\beta}u^{\beta}=0.
\end{equation}

The non-zero components of $\mathcal{E}_{\alpha\beta}$ \cite{grtensor}, written in terms of the mass-function and the shear scalar, are found to be as follows :

\begin{align}\label{Weyl11}
\mathcal{E}_{11} &=-\frac{B^{2}}{3}\left[\frac{2m}{C^{3}}-\frac{Q^{2}}{C^{4}}+\frac{1}{A^{2}}\left(\frac{\ddot{B}}{B}-\frac{\ddot{C}}{C}\right)\right.\nonumber \\
& -\frac{1}{B^{2}}\left(\frac{A''}{A}-\frac{C''}{C}\right)+\frac{B'}{B^{3}}\left(\frac{A'}{A}-\frac{C'}{C}\right)\nonumber \\
& \left.+\frac{3\dot{A}\sigma}{A^{2}}+\frac{C'A'}{AB^{2}C}-\frac{\dot{C}\dot{B}}{A^{2}BC}\right]
\end{align}

\begin{align}\label{Weyl22}
\mathcal{E}_{22} &=\frac{C^{2}}{6}\left[\frac{2m}{C^{3}}-\frac{Q^{2}}{C^{4}}+\frac{1}{A^{2}}\left(\frac{\ddot{B}}{B}-\frac{\ddot{C}}{C}\right)\right. \nonumber \\
& -\frac{1}{B^{2}}\left(\frac{A''}{A}-\frac{C''}{C}\right)+\frac{B'}{B^{3}}\left(\frac{A'}{A}-\frac{C'}{C}\right)\nonumber \\
& \left.+\frac{3\dot{A}\sigma}{A^{2}}+\frac{C'A'}{AB^{2}C}-\frac{\dot{C}\dot{B}}{A^{2}BC}\right]
\end{align}

\begin{align}\label{Weyl33}
\mathcal{E}_{33} &=\frac{C^{2}\sin^{2}\theta}{6}\left[\frac{2m}{C^{3}}-\frac{Q^{2}}{C^{4}}+\frac{1}{A^{2}}\left(\frac{\ddot{B}}{B}-\frac{\ddot{C}}{C}\right)\right. \nonumber \\
& -\frac{1}{B^{2}}\left(\frac{A''}{A}-\frac{C''}{C}\right)+\frac{B'}{B^{3}}\left(\frac{A'}{A}-\frac{C'}{C}\right)\nonumber \\
& \left.+\frac{3\dot{A}\sigma}{A^{2}}+\frac{C'A'}{AB^{2}C}-\frac{\dot{C}\dot{B}}{A^{2}BC}\right]
\end{align}

The Weyl Scalar $\mathcal{E}$ is given by
\begin{align}\label{weyl}
\mathcal{E} &=-\frac{1}{2}\left[\frac{2m}{C^{3}}-\frac{Q^{2}}{C^{4}}+\frac{1}{A^{2}}\left(\frac{\ddot{B}}{B}-\frac{\ddot{C}}{C}\right)\right. \nonumber \\
& -\frac{1}{B^{2}}\left(\frac{A''}{A}-\frac{C''}{C}\right)+\frac{B'}{B^{3}}\left(\frac{A'}{A}-\frac{C'}{C}\right)\nonumber \\
& \left.+\frac{3\dot{A}\sigma}{A^{2}}+\frac{C'A'}{AB^{2}C}-\frac{\dot{C}\dot{B}}{A^{2}BC}\right]
\end{align}

which, with the help of equations \eqref{xtf} and \eqref{ytf}, can be expressed as
\begin{equation}\label{weyl1}
\mathcal{E}=\frac{1}{2}\left(Y_{TF}-X_{TF}\right)
\end{equation}

On evaluating $Z=\sqrt{Z^{\alpha\beta}Z_{\alpha\beta}}$ , from equation \eqref{zab}, we find that
\begin{equation}\label{z}
Z=\left[\left(\frac{2m}{C^{3}}-\frac{Q^{2}}{C^{4}}\right)^{2}+2\left(\frac{2m}{C^{3}}-\frac{Q^{2}}{C^{4}}-X_{TF}\right)^{2}\right]^{\frac{1}{2}}
\end{equation}

\section{The Field Equations for the Interior Spacetime}

We consider our matter Lagrangian to be $-\rho$ , and using this
in our $f(R,T)$ field equation, the effective energy-momentum tensor
is given by
\begin{align}\label{effEMT}
T_{\mu\nu}^{eff} &=\frac{1}{f_{R}}\left[\left(1+f_{T}\right)T_{\mu\nu}+\rho g_{\mu\nu}f_{T}\right. \nonumber \\
& \left.+\frac{\left(f-Rf_{R}\right)}{2}g_{\mu\nu}-D_{\mu\nu}+8\pi E_{\mu\nu}\right]
\end{align}

where $D_{\mu\nu}=\left(g_{\mu\nu}\Box-\nabla_{\mu}\nabla_{\nu}\right)f_{R}$
represents the dark energy source terms. The field equations are found
to be

\begin{equation}\label{G00}
G_{00}=\frac{A^{2}}{f_{R}}\left[\rho+\epsilon+\epsilon f_{T}-\frac{\left(f-Rf_{R}\right)}{2}-\frac{D_{00}}{A^{2}}+\frac{3Q^{2}}{C^{4}}\right]
\end{equation}

\begin{equation}\label{G01}
G_{01}=\frac{AB}{f_{R}}\left[-\left(1+f_{T}\right)\left(q+\epsilon\right)-\frac{D_{01}}{AB}\right]
\end{equation}

\begin{align}\label{G11}
G_{11} &=\frac{B^{2}}{f_{R}}\left[\left(1+f_{T}\right)\left(p_{r}+\epsilon+4\eta\sigma\right)+\rho f_{T}\right. \nonumber \\
& \left.+\frac{\left(f-Rf_{R}\right)}{2}-\frac{D_{11}}{B^{2}}-\frac{3Q^{2}}{C^{4}}\right]
\end{align}

\begin{align}\label{G22}
G_{22} &=\frac{C^{2}}{f_{R}}\left[\left(1+f_{T}\right)\left(p_{\perp}-2\eta\sigma\right)+\rho f_{T}\right. \nonumber \\
& \left.+\frac{\left(f-Rf_{R}\right)}{2}-\frac{D_{22}}{C^{2}}-\frac{Q^{2}}{C^{4}}\right]
\end{align}

Here the coefficient of shear viscosity, $\eta$, is positive. But the shear scalar is a negative quantity, as given by \eqref{shear}. As it can be seen from \eqref{G11} and \eqref{G22}, the shear causes a decrease in the radial pressure, but contributes positively to the tangential pressure, which is natural, as shear viscosity arises out of different fluid layers moving against one another in the tangential direction. Further, for the situation where bulk viscosity might be present, the bulk viscosity will bring a negative contribution, opposing both the radial and tangential pressures. It is possible that for certain values, it might make the effective pressure negative, which would imply a repulsive gravity or dark energy effect. Hence bulk viscosity may possibly account for the accelerated expansion of the universe in this manner \cite{GagnonJCAP2011}.
The dark source terms are given by
\begin{equation}\label{D00}
D_{00}=\dot{f_{R}}\left(\frac{2\dot{C}}{C}+\frac{\dot{B}}{B}\right)-\frac{A^{2}}{B^{2}}f_{R}''-\frac{f_{R}'A^{2}}{B^{2}}\left(\frac{2C'}{C}-\frac{B'}{B}\right)
\end{equation}

\begin{equation}\label{D01}
D_{01}=\frac{f_{R}'\dot{B}}{B}+\frac{\dot{f_{R}}A'}{A}-\dot{f_{R}}'
\end{equation}

\begin{equation}\label{D11}
D_{11}=f_{R}'\left(\frac{A'}{A}+\frac{2C'}{C}\right)-\ddot{f_{R}}\frac{B^{2}}{A^{2}}-\frac{\dot{f_{R}}B^{2}}{A^{2}}\left(\frac{2\dot{C}}{C}-\frac{\dot{A}}{A}\right)
\end{equation}

\begin{align}\label{D22}
D_{22} &=-\frac{\dot{f_{R}}C^{2}}{A^{2}}\left(\frac{\dot{C}}{C}+\frac{\dot{B}}{B}-\frac{\dot{A}}{A}\right)-\ddot{f_{R}}\frac{C^{2}}{A^{2}}\nonumber \\
& +\frac{C^{2}}{B^{2}}f_{R}''+\frac{f_{R}'C^{2}}{B^{2}}\left(\frac{C'}{C}-\frac{B'}{B}+\frac{A'}{A}\right)
\end{align}

Likewise in \cite{Herrera_et_al2009, Hussain2017, BhattiEPJC2021, BhattiCJP2021, BhattiPhysScr2021}, the following quantities are now introduced :

\begin{equation}\label{barredquantities}
\bar{\rho}=\rho+\epsilon
, \;\;\bar{q}=q+\epsilon , \;\;\bar{p_{r}}=p_{r}+\epsilon , \;\;\bar{\Pi}=\bar{p_{r}}-p_{\perp}
.
\end{equation}
The temporal variation and spatial variation of a quantity are represented
by the following operators defined as
\begin{equation}\label{DT}
D_{T}\equiv\frac{1}{A}\frac{\partial}{\partial t}
\end{equation}

and
\begin{equation}\label{DC}
D_{C}\equiv\frac{1}{C'}\frac{\partial}{\partial r}
\end{equation}

The collapse velocity $U$ is given by $U=\frac{\dot{C}}{A}$ which
must be negative for collapse to occur. We also define $H=\frac{C'}{B}$.
Utilising all these definitions, we find the temporal variation of
the function $m$ to be
\begin{align}\label{DTm}
D_{T}m &=-\frac{C^{2}}{2f_{R}}\left[U\left\{ \left(1+f_{T}\right)\left(\bar{p_{r}}+4\eta\sigma\right)+\rho f_{T}\right.\right. \nonumber\\
& \left.\left.+\frac{\left(f-Rf_{R}\right)}{2}-\frac{D_{11}}{B^{2}}-\frac{3Q^{2}}{C^{4}}\right\}\right. \nonumber \\
& \left.+H\left\{ \bar{q}\left(1+f_{T}\right)+\frac{D_{01}}{AB}\right\} \right]-\frac{UQ^{2}}{C^{2}}
\end{align}

while the spatial variation is given by
\begin{align}\label{DCm}
D_{C}m &=\frac{C^{2}}{2f_{R}}\left[\bar{\rho}+\epsilon f_{T}-\frac{\left(f-Rf_{R}\right)}{2}-\frac{D_{00}}{A^{2}}+\frac{3Q^{2}}{C^{4}}\right.\nonumber\\
& \left.+\frac{U}{H}\left\{ \bar{q}\left(1+f_{T}\right)+\frac{D_{01}}{AB}\right\} \right]+D_{C}\left(\frac{Q^{2}}{2C}\right)
\end{align}
Hence, the function $m$ can be expressed as
\begin{align}\label{massfn1}
m &=\int^{r}_{0}\frac{C^{2}}{2f_{R}}\left[\bar{\rho}+\epsilon f_{T}-\frac{\left(f-Rf_{R}\right)}{2}-\frac{D_{00}}{A^{2}}+\frac{3Q^{2}}{C^{4}}\right.\nonumber\\
& \left.+\frac{U}{H}\left\{ \bar{q}\left(1+f_{T}\right)+\frac{D_{01}}{AB}\right\} \right]C'dr+\frac{Q^{2}}{2C}.
\end{align}
As we can see, the mass-energy content of the collapsing matter, which, by virtue of the field equations, is described by $m$, is increased by the presence of electric charge.

\section{The Relation of the Matter Variables to the Structure Scalars}

The various physical properties of the collapsing matter, which involve the energy density, pressure anisotropy, expansion and shear, can be related to the structure scalars formed out of the Riemann tensor which is a purely geometrical quantity.
Utilising the field equations and the Weyl scalar, the structure scalars
can be expressed as

\begin{equation}\label{xt1}
X_{T}=\frac{1}{f_{R}}\left[\bar{\rho}+\epsilon f_{T}-\frac{\left(f-Rf_{R}\right)}{2}-\frac{D_{00}}{A^{2}}+\frac{3Q^{2}}{C^{4}}\right]
\end{equation}

\begin{equation}\label{xtf1}
X_{TF}=-\mathcal{E}-\frac{1}{2f_{R}}\left[\left(1+f_{T}\right)\left(\bar{\Pi}+6\eta\sigma\right)-\frac{D_{11}}{B^{2}}+\frac{D_{22}}{C^{2}}-\frac{2Q^{2}}{C^{4}}\right]
\end{equation}

\begin{align}\label{yt1}
Y_{T} &=\frac{1}{2f_{R}}\left[3\rho f_{T}+\bar{\rho}+\epsilon f_{T}+\left(1+f_{T}\right)\left(\bar{p_{r}}+2p_{\perp}\right)\right.\nonumber \\
& \left.+f-Rf_{R}-\frac{D_{00}}{A^{2}}-\frac{D_{11}}{B^{2}}-\frac{2D_{22}}{C^{2}}-\frac{2Q^{2}}{C^{4}}\right]
\end{align}

\begin{equation}\label{ytf1}
Y_{TF}=\mathcal{\mathcal{E}}-\frac{1}{2f_{R}}\left[\left(1+f_{T}\right)\left(\bar{\Pi}+6\eta\sigma\right)-\frac{D_{11}}{B^{2}}+\frac{D_{22}}{C^{2}}-\frac{2Q^{2}}{C^{4}}\right]
\end{equation}

It can be seen that the presence of electric charge causes an increment in the structure scalars $X_{T}$ and $Y_{TF}$, and a decrement in $X_{TF}$ and $Y_{T}$.

Now the matter variables are combined in the following manner to obtain quantites which can be termed as ``effective energy density", ``effective radial pressure", ``effective tangential pressure" and ``effective pressure anistropy", and are denoted with the superscript $\mathscr{P}$ :
\begin{equation}\label{rhoP}
\rho^{\mathscr{P}}=\bar{\rho}+\epsilon f_{T}-\frac{D_{00}}{A^{2}}
\end{equation}

\begin{equation}\label{PrP}
P_{r}^{\mathscr{P}}=\bar{p_{r}}+4\eta\sigma-\frac{D_{11}}{B^{2}}
\end{equation}

\begin{equation}\label{PperpP}
P_{\perp}^{\mathscr{P}}=p_{\perp}-2\eta\sigma-\frac{D_{22}}{C^{2}}
\end{equation}

\begin{equation}\label{PiP}
\Pi^{\mathscr{P}}=P_{r}^{\mathscr{P}}-P_{\perp}^{\mathscr{P}}=\bar{\Pi}+6\eta\sigma-\frac{D_{11}}{B^{2}}+\frac{D_{22}}{C^{2}}
\end{equation}

Hence we have the Weyl scalar expressed in terms of the physical matter variables in the following manner,
\begin{align}\label{weyl2}
\mathcal{E} &=\frac{1}{2f_{R}}\left[\rho^{\mathscr{P}}-\frac{\left(f-Rf_{R}\right)}{2}+\frac{5Q^{2}}{C^{4}}+\frac{D_{11}}{B^{2}}-\frac{D_{22}}{C^{2}}\right.\nonumber \\
& \left.-\left(1+f_{T}\right)\left(\bar{\Pi}+6\eta\sigma\right)\right]-I
\end{align}

where,
\begin{align}\label{I}
I &=\frac{3}{2C^{3}}\int_{0}^{r}\frac{C^{2}}{f_{R}}\left[\rho^{\mathscr{P}}-\frac{\left(f-Rf_{R}\right)}{2}+\frac{3Q^{2}}{C^{4}}\right.\nonumber \\
& \left.+\frac{U}{H}\left\{ \hat{q}+\frac{\psi_{q}}{AB}\right\} \right]C'dr
\end{align}

with $\hat{q}=\bar{q}\left(1+f_{T}\right)$ and $\psi_{q}=D_{01}$. Now, in terms of the effective matter variables, our structure scalars become
\begin{equation}\label{xt2}
X_{T}=\frac{1}{f_{R}}\left[\rho^{\mathscr{P}}-\frac{\left(f-Rf_{R}\right)}{2}+\frac{3Q^{2}}{C^{4}}\right]
\end{equation}

\begin{equation}\label{xtf2}
X_{TF}=-\frac{1}{2f_{R}}\left[\rho^{\mathscr{P}}-\frac{\left(f-Rf_{R}\right)}{2}+\frac{3Q^{2}}{C^{4}}\right]+I
\end{equation}

\begin{align}\label{yt2}
Y_{T} &=\frac{1}{2f_{R}}\left[3\rho f_{T}+\rho^{\mathscr{P}}+\left(f-Rf_{R}\right)+\frac{D_{11}}{B^{2}}+\frac{2D_{22}}{C^{2}}\right.\nonumber\\
& \left.+\left(1+f_{T}\right)\left(3\bar{p_{r}}-2\bar{\Pi}\right)-\frac{2Q^{2}}{C^{4}}\right]
\end{align}

\begin{align}\label{ytf2}
Y_{TF} &=\frac{1}{2f_{R}}\left[\rho^{\mathscr{P}}-\frac{\left(f-Rf_{R}\right)}{2}-\frac{2D_{11}}{B^{2}}+\frac{2D_{22}}{C^{2}}\right.\nonumber\\
& \left.+\frac{7Q^{2}}{C^{4}}-2\left(1+f_{T}\right)\left(\bar{\Pi}+6\eta\sigma\right)\right]-I
\end{align}

From the expressions of the structure scalars in terms of the metric
coefficients, and the Raychaudhuri equation \cite{Raychaudhuri1955}, we see that two of our
structure scalars can be expressed in the following manner :

\begin{equation}\label{ytmodified}
-Y_{T}=\frac{\Theta^{2}}{3}+6\sigma^{2}+u^{\alpha}\Theta_{;\alpha}-a_{;\alpha}^{\alpha}
\end{equation}

and
\begin{equation}\label{ytfmodified}
Y_{TF}=a^{2}+\chi^{\alpha}a_{;\alpha}-\frac{aC'}{BC}+2\Theta\sigma-3u^{\alpha}\sigma_{;\alpha}-3\sigma^{2}
\end{equation}

where, $a^{\alpha}$ is the acceleration vector.
The first of these expressions show how the expansion of the collapsing spherical star evolves during the collapse, which is governed by the structure scalar $Y_{T}$. The second expression describes the shear evolution of the star and is controlled by the structure scalar $Y_{TF}$.

Utilising the expressions for the shear scalar $\sigma$, the expansion
scalar $\Theta$, $X_{TF}$ , and taking the radial derivative of
the expression for $X_{TF}$ , we get

\begin{align}\label{xtfmodified}
\left[X_{TF}+\frac{\rho_{eff}}{2f_{R}}\right]'& =-\frac{3C'}{C}X_{TF}+\frac{\left(\Theta+3\sigma\right)}{2f_{R}}\left(\hat{q}B+\frac{\psi_{q}}{A}\right)\nonumber\\
&+\frac{9C'}{C^{4}}\left(m-\frac{Q^2}{2C}\right)
\end{align}

with
\begin{equation}\label{rhoeff}
\rho_{eff}=\rho^{\mathscr{P}}-\frac{\left(f-Rf_{R}\right)}{2}+\frac{3Q^{2}}{C^{4}}.
\end{equation}
Equation \eqref{xtfmodified} can also be rewritten in terms of $X_{TF}$ and $Z$ as

\begin{align}\label{xtfmodifiedwithz}
\left[X_{TF}+\frac{\rho_{eff}}{2f_{R}}\right]'&=-\frac{3C'}{C}X_{TF}+\frac{\left(\Theta+3\sigma\right)}{2f_{R}}\left(\hat{q}B+\frac{\psi_{q}}{A}\right)\nonumber\\
&+\frac{9C'}{2C}\left[Z^{2}-2\left(\frac{2m}{C^{3}}-\frac{Q^{2}}{C^{4}}-X_{TF}\right)^{2}\right]^{\frac{1}{2}}
\end{align}

This expression contains information as to the effect of $X_{TF}$ as well as $Z$ on the energy-density inhomogeneity of the collapsing model, and relates the tidal forces and the heat flux, expansion scalar and the shear scalar.
Hence $X_{T}$ can also be expressed as
\begin{equation}\label{xtrhoeff}
X_{T}=\frac{\rho_{eff}}{f_{R}}
\end{equation}
As it can be seen,
\begin{itemize}
\item $X_{TF}$ and the mass-function $m$ together along with the charge $Q$ influence the energy density inhomogeneity when there is no dissipation, as can be seen from equation \eqref{xtfmodified}. The structure scalar $Z$ is connected to $X_{TF}$, by virtue of equation \eqref{z}. This enables us to see how $Z$ affects the heat dissipation, as well as the energy density inhomogeneity, by a recasting of equation \eqref{xtfmodified} into equation \eqref{xtfmodifiedwithz}.
\item The evolution of the expansion scalar is controlled by $Y_{T}$, where the effect of charge is already included in the expression for $Y_{T}$, given by equation \eqref{yt2}.
\item The evolution of the shear scalar is controlled by $Y_{TF}$, where the effect of charge is present in the expression for $Y_{TF}$, given by equation \eqref{ytf2}.
\item The effective homogeneous energy density is influenced by $X_{T}$.
\item The pressure anisotropy is also influenced by both $Y_{T}$ and $Y_{TF}$.
\item The electric charge $Q$ appears in the expressions for all four of the structure scalars, causing an increase in $X_{T}$ and $Y_{TF}$, and a decrease in $X_{TF}$ and $Y_{T}$.
\end{itemize}

\subsection{Case of Constant R and T conditions for dust ball}

By $R=constant=\tilde{R}$ and $T=constant=\tilde{T}$, we have
\begin{equation}\label{xtconstant}
X_{T}=constant=\tilde{X}_{T}
\end{equation}

\begin{equation}\label{ytconstant}
Y_{T}=constant=\tilde{Y}_{T}
\end{equation}

where $\tilde{R}$ and $\tilde{T}$  denote the values of the Ricci tensor and the Trace of the enegy-momentum tensor, at constant $R$ and $T$ conditions.

For the case of dust, except the energy density, all the other physical parameters, i.e., radial and tangential pressures, shear, free-streaming radiation, heat flux and charge will vanish, and $\bar{\rho}$ becomes the same as $\rho$. In that case, we have the structure scalars related to the matter in the following forms :

\begin{equation}\label{xt3}
\tilde{X}_{T}=\frac{1}{f_{R}}\left[\rho-\frac{\left(f-Rf_{R}\right)}{2}-\frac{D_{00}}{A^{2}}\right]
\end{equation}

\begin{equation}\label{yt3}
\tilde{Y}_{T}=\frac{1}{2f_{R}}\left[\left(1+f_{T}\right)\left(2\rho+T\right)+2\rho f_{T}+f-Rf_{R}\right]
\end{equation}

\begin{equation}\label{xtf3}
\tilde{X}_{TF}=-\mathcal{E}
\end{equation}

\begin{equation}\label{ytf3}
\tilde{Y}_{TF}=\mathcal{E}
\end{equation}

It can be seen that the sum of $X_{TF}$ and $Y_{TF}$ vanishes in this special case.

\subsection{Choosing a linear form of the $f(R,T)$ function}

Choosing a linear form of the $f(R,T)$ function, where $f(R,T)=R+\lambda T$ gives rise to a power-law type of scale factor. In this case, $f_{R}=1$, $f_{T}=\lambda$, and $D_{\mu\nu}=0$ . Utilising this functional form, the relations between the matter variables and the structure scalars take the following form :

\begin{equation}\label{xt4}
X_{T}=\bar{\rho}+\epsilon\lambda -\frac{\lambda T}{2}+\frac{3Q^{2}}{C^{4}}
\end{equation}

\begin{equation}\label{xtf4}
X_{TF}=-\mathcal{E}-\frac{1}{2}\left[\left(1+\lambda\right)\left(\bar{\Pi}+6\eta\sigma\right)-\frac{2Q^{2}}{C^{4}}\right]
\end{equation}

\begin{equation}\label{yt4}
Y_{T}=\frac{1}{2}\left[3\rho\lambda+\bar{\rho}+\epsilon\lambda+\left(1+\lambda\right)\left(\bar{p_{r}}+2p_{\perp}\right)+\lambda T-\frac{2Q^{2}}{C^{4}}\right]
\end{equation}

\begin{equation}\label{ytf4}
Y_{TF}=\mathcal{\mathcal{E}}-\frac{1}{2}\left[\left(1+\lambda\right)\left(\bar{\Pi}+6\eta\sigma\right)-\frac{2Q^{2}}{C^{4}}\right]
\end{equation}

The ``effective" matter variables after being combined together, now become :

\begin{equation}\label{rhoP1}
\rho^{\mathscr{P}}=\bar{\rho}+\epsilon\lambda
\end{equation}

\begin{equation}\label{PrP1}
P_{r}^{\mathscr{P}}=\bar{p_{r}}+4\eta\sigma
\end{equation}

\begin{equation}\label{PperpP1}
P_{\perp}^{\mathscr{P}}=p_{\perp}-2\eta\sigma
\end{equation}

\begin{equation}\label{PiP1}
\Pi^{\mathscr{P}}=P_{r}^{\mathscr{P}}-P_{\perp}^{\mathscr{P}}=\bar{\Pi}+6\eta\sigma
\end{equation}

In terms of the effective matter variables, the structure scalars can be expressed as :

\begin{equation}\label{xt5}
X_{T}=\rho^{\mathscr{P}} -\frac{\lambda T}{2}+\frac{3Q^{2}}{C^{4}}
\end{equation}

\begin{equation}\label{xtf5}
X_{TF}=-\mathcal{E}-\frac{1}{2}\left[\left(1+\lambda\right)\Pi^{\mathscr{P}}-\frac{2Q^{2}}{C^{4}}\right]
\end{equation}

\begin{equation}\label{yt5}
Y_{T}=\frac{1}{2}\left[3\rho\lambda+\rho^{\mathscr{P}}+\left(1+\lambda\right)\left(P_{r}^{\mathscr{P}}+2P_{\perp}^{\mathscr{P}}\right)+\lambda T-\frac{2Q^{2}}{C^{4}}\right]
\end{equation}

\begin{equation}\label{ytf5}
Y_{TF}=\mathcal{\mathcal{E}}-\frac{1}{2}\left[\left(1+\lambda\right)\Pi^{\mathscr{P}}-\frac{2Q^{2}}{C^{4}}\right]
\end{equation}

\subsection{Complexity Factor}

Now we proceed to discuss the effect of complexity on the progress of collapse of the matter configuration under our consideration. Utilising equations \eqref{xt}-\eqref{ytf}, and \eqref{weyl}, the Weyl scalar can be expressed as :

\begin{equation}\label{recastweyl}
\mathcal{E}=\frac{Y_{TF}}{2}+\frac{1}{4C^{3}}\int C^{3}\left(\frac{G_{00}}{A^{2}}\right)'dr+\frac{3}{4C^{3}}\int\frac{\dot{C}C^{2}G_{01}}{A^{2}}dr
\end{equation}
From equation \eqref{G00}, it can be seen that the quantity $\frac{G_{00}}{A^{2}}$ in the second term of \eqref{recastweyl} is proportional to the interior energy density $\rho$. Hence its radial derivative must be proportional to $\rho'$ which is the inhomgeneity of energy density. Hence the Weyl scalar also depends on the inhomogeneity of the energy density.
Utilising equation \eqref{recastweyl}, equation \eqref{ytf1} can be rewritten as

\begin{align}\label{ytfcomplexity}
Y_{TF} &=\frac{1}{2C^{3}}\int C^{3}\left(\frac{G_{00}}{A^{2}}\right)'dr+\frac{3}{2C^{3}}\int\frac{\dot{C}C^{2}G_{01}}{A^{2}}dr\nonumber \\
& -\frac{1}{f_{R}}\left[\left(1+f_{T}\right)\left(\bar{\Pi}+6\eta\sigma\right)-\frac{D_{11}}{B^{2}}+\frac{D_{22}}{C^{2}}-\frac{2Q^{2}}{C^{4}}\right]
\end{align}

As it can be seen, from equations \eqref{G00} and \eqref{G01}, in case of GR, and in absence of all dissipative and electromagnetic effects, for isotropic pressure and homogeneous energy density, that is, for zero complexity, the structure scalar $Y_{TF}$ vanishes. Hence this particular structure scalar is a reasonable choice for describing the complexity of the collapsing system, and is known as the complexity factor. It is worth mentioning that a vanishing complexity can also arise from the condition $Y_{TF}=0$, which gives
\begin{align}\label{complexitycondition}
\frac{1}{2C^{3}}\int C^{3}\left(\frac{G_{00}}{A^{2}}\right)'dr+\frac{3}{2C^{3}}\int\frac{\dot{C}C^{2}G_{01}}{A^{2}}dr\nonumber \\
-\frac{1}{f_{R}}\left[\left(1+f_{T}\right)\left(\bar{\Pi}+6\eta\sigma\right)-\frac{D_{11}}{B^{2}}+\frac{D_{22}}{C^{2}}-\frac{2Q^{2}}{C^{4}}\right] &=0.
\end{align}

\section{Exterior Spacetime}
We have considered the exterior spacetime to described by the generalized Vaidya metric of outgoing radiation, whose line element is given by :
\begin{equation}\label{genvaidya}
ds_{+}^{2}=-\left(1-\frac{2M(v,Y)}{Y}\right)dv^{2}-2dvdY+Y^{2}d\Omega^{2},
\end{equation}
where $d\Omega^{2}=d\theta^{2}+\sin^{2}\theta d\phi^{2}$. Here, $v$ is the retarded null coordinate, and $Y$ is the radial coordinate for the exterior spacetime. The mass-energy content inside a radius $Y$ at a time $v$ is described by the function $M(v,Y)$.

The exterior energy-momentum tensor is assumed to be a composite matter which is a combination of type-I and type-II fluids \cite{WangGRG1999}, and is expressed as follows:
\begin{equation}
T_{\mu\nu}^{+}=\mu l_{\mu}l_{\nu}+\left(\rho+P\right)\left(l_{\mu}n_{\nu}+l_{\nu}n_{\mu}\right)+Pg_{\mu\nu}^{+},
\end{equation}
where the first term represents a type-I fluid of density $\mu$, representing null-like matter or radiation photons, and the remaining terms together represent a type-II fluid, with $\rho$ being its density and $P$ being its isotropic pressure, which represents a perfect fluid component which can represent timelike/massive particles. Types of matter are further discussed in \cite{HawkingEllis}. Here,
\begin{equation}
l_{\mu}=\delta_{\mu}^{0},
\end{equation}
\begin{equation}
\textrm{and} \qquad n_{\mu}=\frac{1}{2}\left(1-\frac{2M(v,Y)}{Y}\right)\delta_{\mu}^{0}-\delta_{\mu}^{1} .
\end{equation}
The trace of the exterior energy-momentum tensor is
\begin{equation}\label{Traceexterior}
T^{+}=6P+2\rho .
\end{equation}
The $f(R,T)$ field equations for the exterior is given by :
\begin{equation}\label{fRTfe}
R_{\mu\nu}=\frac{1}{f_{R}}\left[\left(1+f_{T}\right)T^{m}_{\mu\nu}-L_{m_{ext}}g_{\mu\nu}f_{T} +\frac{1}{2}g_{\mu\nu}f-D_{\mu\nu}\right],
\end{equation}
which gives us, using \eqref{genvaidya}
\begin{align}
\label{R00+}R_{00} & =\frac{1}{Y^{2}}\left(\left(\frac{\partial^{2}M}{\partial Y^{2}}\right)\left(2M-Y\right)-2\frac{\partial M}{\partial v}\right) \nonumber \\
& =\frac{1}{f_{R}}\left[\left(1+f_{T}\right)\left(\mu+\rho_{ext}\left(1-\frac{2M}{Y}\right)\right)\right.\nonumber\\
& \left.+\left(1-\frac{2M}{Y}\right)L_{m_{ext}}f_{T}-\frac{1}{2}\left(1-\frac{2M}{Y}\right)f-D_{00}^{+}\right],\\
\label{R11+}R_{11} & =0=\frac{-D_{11}^{+}}{f_{R}},\\
\label{R22+}R_{22} & =2\frac{\partial M}{\partial Y}\nonumber\\
& =\frac{1}{f_{R}}\left[\left(1+f_{T}\right)PY^{2}-L_{m_{ext}}f_{T}Y^{2}+\frac{Y^{2}}{2}f-D_{22}^{+}\right],\\
\label{R33+}R_{33} & =R_{22}\sin^{2}\theta,\\
\label{R01+}R_{01} & =-\frac{1}{Y}\frac{\partial^{2}M}{\partial Y^{2}}\nonumber\\
& =\frac{1}{f_{R}}\left[\left(1+f_{T}\right)\left(-\rho_{ext}-2P\right)+L_{m_{ext}}f_{T}-\frac{f}{2}-D_{01}^{+}\right].
\end{align}
where, $L_{m_{ext}}$ is the matter Lagrangian for the exterior spacetime, and a comma indicates a partial derivative.
Here, the exterior dark source terms are given by
\begin{align}
\label{D00+}D_{00}^{+} & =\frac{1}{Y^{3}}\left[f_{R_{,YY}}\left(-4M^{2}Y+4MY^{2}-Y^{3}\right)\right.\nonumber\\
& +f_{R_{,Y}}\left(-2M_{,Y}MY+M_{,Y}Y^{2}\right.\nonumber \\
& \left.-M_{,v}Y^{2}-6M^{2}+7MY-2Y^{2}\right) \nonumber \\
& +f_{R_{,vY}}\left(-4MY^{2}+2Y^{3}\right)-f_{R_{,vv}}Y^{3} \nonumber \\
& \left.+f_{R_{,v}}\left(M_{,Y}Y^{2}-5MY+2Y^{2}\right)\right], \\
\label{D11+}D_{11}^{+} & =-f_{R_{,YY}}, \\
\label{D22+}D_{22}^{+} & =Y\left[f_{R_{,YY}}\left(-2M+Y\right)+f_{R_{,Y}}\left(-2M_{,Y}+1\right)\right.\nonumber\\
& \left.-2f_{R_{,Yv}}Y-f_{R_{,v}}\right], \\
\label{D01+}D_{01}^{+} & =\frac{f_{R_{,YY}}}{Y^{2}}\left(2MY-Y^{2}\right)+f_{R_{,Yv}}+\frac{2}{Y}f_{R_{,v}}\nonumber \\
& +\frac{f_{R_{,Y}}}{Y^{2}}\left(M_{,Y}Y+3M-2Y\right).
\end{align}
Using \eqref{R11+} and \eqref{D11+}, we have
\begin{equation}\label{fRTconstraint}
f_{R_{,YY}}=0,
\end{equation}
which simplifies the dark source terms in the following manner :
\begin{align}
D_{00}^{+} & =\frac{1}{Y^{3}}\left[f_{R_{,Y}}\left(-2M_{,Y}MY+M_{,Y}Y^{2}-M_{,v}Y^{2}\right.\right.\nonumber\\
& \left.-6M^{2}+7MY-2Y^{2}\right) \nonumber \\
& +f_{R_{,vY}}\left(-4MY^{2}+2Y^{3}\right)-f_{R_{,vv}}Y^{3}\nonumber\\
& \left.+f_{R_{,v}}\left(M_{,Y}Y^{2}-5MY+2Y^{2}\right)\right], \\
D_{11}^{+} & =0, \\
D_{22}^{+} & =Y\left[f_{R_{,Y}}\left(-2M_{,Y}+1\right)-2f_{R_{,Yv}}Y-f_{R_{,v}}\right], \\
D_{01}^{+} & =\frac{f_{R_{,Y}}}{Y^{2}}\left(M_{,Y}Y+3M-2Y\right)+f_{R_{,Yv}}+\frac{2}{Y}f_{R_{,v}}.
\end{align}

\section{The Interior Field Equations}
The matter-Lagrangian for the interior is kept unspecified for this part of the analysis, and denoted by $L_{m_{int}}$. The field equations for the interior spacetime, re-expressed in terms of the components of the Ricci tensor, are found
to be
\begin{align}\label{R00-}
R_{00}^{-} &=\frac{A^{2}}{f_{R}}\left[\left(1+f_{T}\right)\left(\rho+\epsilon\right)+L_{m_{int}} f_{T}\right.\nonumber\\
& \left.-\frac{f}{2}-\frac{D_{00}^{-}}{A^{2}}+\frac{3Q^{2}}{C^{4}}\right]
\end{align}

\begin{equation}\label{R01-}
R_{01}^{-}=\frac{AB}{f_{R}}\left[-\left(1+f_{T}\right)\left(q+\epsilon\right)-\frac{D_{01}^{-}}{AB}\right]
\end{equation}

\begin{align}\label{R11-}
R_{11}^{-} &=\frac{B^{2}}{f_{R}}\left[\left(1+f_{T}\right)\left(p_{r}+\epsilon+4\eta\sigma\right)-L_{m_{int}} f_{T}\right.\nonumber\\
& \left.+\frac{f}{2}-\frac{D_{11}^{-}}{B^{2}}-\frac{3Q^{2}}{C^{4}}\right]
\end{align}

\begin{align}\label{R22-}
R_{22}^{-} &=\frac{C^{2}}{f_{R}}\left[\left(1+f_{T}\right)\left(p_{\perp}-2\eta\sigma\right)-L_{m_{int}} f_{T}\right.\nonumber\\
& \left.+\frac{f}{2}-\frac{D_{22}^{-}}{C^{2}}-\frac{Q^{2}}{C^{4}}\right]
\end{align}

The dark source terms are given by
\begin{equation}
D_{00}^{-}=\dot{f_{R}}\left(\frac{2\dot{C}}{C}+\frac{\dot{B}}{B}\right)-\frac{A^{2}}{B^{2}}f_{R}''-\frac{f_{R}'A^{2}}{B^{2}}\left(\frac{2C'}{C}-\frac{B'}{B}\right)
\end{equation}
\begin{equation}
D_{01}^{-}=\frac{f_{R}'\dot{B}}{B}+\frac{\dot{f_{R}}A'}{A}-\dot{f_{R}}'
\end{equation}

\begin{equation}
D_{11}^{-}=f_{R}'\left(\frac{A'}{A}+\frac{2C'}{C}\right)-\ddot{f_{R}}\frac{B^{2}}{A^{2}}-\frac{\dot{f_{R}}B^{2}}{A^{2}}\left(\frac{2\dot{C}}{C}-\frac{\dot{A}}{A}\right)
\end{equation}

\begin{align}
D_{22}^{-} &=-\frac{\dot{f_{R}}C^{2}}{A^{2}}\left(\frac{\dot{C}}{C}+\frac{\dot{B}}{B}-\frac{\dot{A}}{A}\right)-\ddot{f_{R}}\frac{C^{2}}{A^{2}}\nonumber\\
& +\frac{C^{2}}{B^{2}}f_{R}''+\frac{f_{R}'C^{2}}{B^{2}}\left(\frac{C'}{C}-\frac{B'}{B}+\frac{A'}{A}\right)
\end{align}

The Trace of the interior energy-momentum tensor is given by
\begin{equation}\label{traceint}
T^{-}=-\rho+p_{r}+2p_{\perp} .
\end{equation}
For the case where bulk viscosity is present, the bulk viscosity term will be present in the Trace. It will oppose and reduce the net effective pressure. The effective pressure may also become negative in that case, mimicking the dark energy phase of the evolution of the universe when it undergoes late-time accelerated expansion. So, the bulk viscosity term can be vital in explaining this accelerated expanding phase \cite{GagnonJCAP2011}.

\section{Junction Conditions}

The junction conditions, or the continuity of quantities such as the spacetime metric, the trace part and tracefree part of the extrinsic curvature tensor, Ricci scalar, trace of the energy-momentum tensor, and their derivatives with respect to the corresponding interior and exterior coordinates, are essential to the study of gravitational collapse. In GR, only the spacetime metric and the extrinsic curvature tensor components are required to be continuous for a smooth matching across the boundary of the collapsing matter \cite{Darmois, Israel}. The additional quantities are required to be matched across the boundary when one moves to $f(R,T)$ gravity from GR. The junction conditions for $f(R,T)$ gravity were first presented by Rosa \cite{Rosa2021}.
The 3D timelike hypersurface separating the interior and the exterior spacetimes, and forming the boundary of the collapsing matter, is given by
\begin{equation}\label{hyper}
ds_{\Sigma}^{2}=-d\tau^{2}+\mathcal{R}(\tau)^{2}\left(d\theta^{2}+\sin^{2}\theta d\phi^{2}\right),
\end{equation}
where, $\tau$, $\theta$ and $\phi$ are the hypersurface coordinates, later denoted by $\xi^{i}$ collectively.
The junction conditions for a smooth matching in $f(R,T)$ gravity given by Rosa \cite{Rosa2021} are provided as
\begin{align}
\label{jc1}\left[g_{\mu\nu}\right]_{-}^{+} & =0,\\
\label{jc2}\left[\tilde{K}_{ij}\right]_{-}^{+} & =0,\\
\label{jc3}\left[K\right]_{-}^{+} & =0,\\
\label{jc4}\left[R\right]_{-}^{+} & =0,\\
\label{jc5}\left[T\right]_{-}^{+} & =0,\\
\label{jc6}\left[\partial_{\mu}T\right]_{-}^{+} & =0,\\
\label{jc7}\left[\partial_{\mu}R\right]_{-}^{+} & =0.
\end{align}
Here, $K$ and $\tilde{K}_{ij}$ are the trace part and trace-free parts of the extrinsic curvature tensor $K_{ij}$, whose expression is given by
\begin{equation}\label{Kij}
K_{ij}=-N_{\sigma}\left(\frac{\partial^{2}\psi^{\sigma}}{\partial\xi^{i}\partial\xi^{j}} +\Gamma_{\alpha\beta}^{\sigma}\frac{\partial\psi^{\alpha}}{\partial\xi^{i}}\frac{\partial\psi^{\beta}}{\partial\xi^{j}}\right),
\end{equation}
where $\Gamma_{\alpha\beta}^{\sigma}$ are the Christoffel symbols for the spacetime under consideration, $N_{\sigma}$ is the normal to the hypersurface, and $\psi^{\sigma}$ are the coordinates of the 4D-spacetime.

Using the first three junction conditions, \eqref{jc1}, \eqref{jc2} and \eqref{jc3}, and the field equations, \eqref{R01-} and \eqref{R11-}, we obtain,
\begin{align}\label{boundaryrelation}
-\frac{C}{2}\left[\frac{1}{f_{R}}\left(\left(1+f_{T}\right)\left(p_{r}+4\eta\sigma-q\right)-L_{m_{int}}f_{T}\right.\right.\nonumber\\
\left.\left.+\frac{f-Rf_{R}}{2}-\frac{D_{01}^{-}}{AB}-\frac{D_{11}^{-}}{B^{2}}-\frac{3Q^{2}}{C^{4}}\right)\right]+\frac{Q^{2}}{2C^{3}} \vert_{\Sigma} &=\frac{1}{Y}M_{,Y}\vert_{\Sigma}
\end{align}
which brings out the relation between the radial pressure and heat flux at the boundary, along with the shear viscosity, in presence of the terms arising due to modified gravity. In absence of charge and shear viscosity, for Vaidya metric in GR, this relation reduces to $p_{r}=q$ at the boundary.

From \eqref{jc4} and \eqref{jc5}, it can be seen that for the same functional form of the $f(R,T)$ function in both the interior and the exterior regions, we must have
\begin{equation}\label{fRTcontinuity}
f(R,T)^{-}\vert_{\Sigma}=f(R,T)^{+}\vert_{\Sigma},
\end{equation}
which also leads to the continuity of $f_{R}$ and $f_{T}$ at the boundary. Utilising the continuity of $f(R,T)$, $f_{R}$ and $f_{T}$ across the boundary, and the junction conditions, we find that the matter-Lagrangians for the two spacetimes are related in the following manner :
\begin{align}\label{lagrangiancondition}
4f_{T}\left(L_{m_{int}}-L_{m_{ext}}\right)\vert_{\Sigma} &=\frac{D_{00}^{-}}{A^{2}}-\frac{D_{11}^{-}}{B^{2}}-\frac{2D_{22}^{-}}{C^{2}}\nonumber\\
& -2D_{01}^{+}+\frac{2D_{22}^{+}}{C^{2}}-\frac{8Q^{2}}{C^{4}}\vert_{\Sigma}
\end{align}
In absence of charge, and an $f(R,T)$ function which is linear in both $R$ and $T$, the Lagrangians will match at the boundary, a fact which has been discussed in detail in \cite{GhoshGuha2025}.

Using \eqref{jc6} and \eqref{jc7}, it can be seen that the continuity of the quantities $\partial_{\mu}f$, $\partial_{\mu}f_{R}$ and $\partial_{\mu}f_{T}$ across the boundary are necessary, if the same form of the $f(R,T)$ function is chosen for both the interior and the exterior regions. This leads us to the following two conditions :
\begin{align}\label{tderivative}
-4\frac{\partial}{\partial t}\left(L_{m_{int}}f_{T}\right)+\frac{\partial}{\partial t}\left(\frac{D_{00}^{-}}{A^{2}}-\frac{D_{11}^{-}}{B^{2}}-\frac{2D_{22}^{-}}{C^{2}}\right)-8\frac{\partial}{\partial t}\frac{Q^{2}}{C^{4}}\vert_{\Sigma}\nonumber\\
=-4\frac{\partial}{\partial v}\left(L_{m_{ext}}f_{T}\right)+2\frac{\partial}{\partial v}\left(D_{01}^{+}-\frac{D_{22}^{+}}{Y^{2}}\right)\vert_{\Sigma},
\end{align}
and,
\begin{align}\label{rderivative}
-4\frac{\partial}{\partial r}\left(L_{m_{int}}f_{T}\right)+\frac{\partial}{\partial r}\left(\frac{D_{00}^{-}}{A^{2}}-\frac{D_{11}^{-}}{B^{2}}-\frac{2D_{22}^{-}}{C^{2}}\right)-8\frac{\partial}{\partial r}\frac{Q^{2}}{C^{4}}\vert_{\Sigma}\nonumber\\
=-4\frac{\partial}{\partial Y}\left(L_{m_{ext}}f_{T}\right)+2\frac{\partial}{\partial Y}\left(D_{01}^{+}-\frac{D_{22}^{+}}{Y^{2}}\right)\vert_{\Sigma},
\end{align}
where the effect of charge and modified gravity is clearly brought out, in form of the terms involving $Q$ and the extra curvature terms. Equations \eqref{boundaryrelation}, \eqref{lagrangiancondition},\eqref{tderivative} and \eqref{rderivative} need to be adhered to for a smooth matching at the boundary in order for the collapse to be physically viable. Choosing $f(R,T)=R+2\lambda T$ reproduces the conditions for the matching of the matter-Lagrangians and their derivatives across the boundary which have been presented in equations (47) and (52) of \cite{GhoshGuha2025}.

\section{Energy Conditions in $f(R,T)$ gravity}
Energy conditions are essentially the requirement that the matter under consideration shows a physically realistic behaviour. This rules out any exotic matter contribution. A discussion on energy conditions can be found in \cite{HawkingEllis}. In order to find the energy conditions, following the procedure adopted by Kolassis \cite{Kolassis1988}, and using the effective energy-momentum tensor
$T_{\mu\nu}^{eff}$ instead of the matter energy momentum tensor $T_{\mu\nu}$,
we solve the eigenvalue equation

\begin{equation}\label{eigenvalue}
\left|T_{\mu\nu}^{eff}-\mu g_{\mu\nu}\right|=0
\end{equation}
Let the quantities $\frac{T_{\mu\nu}^{eff}}{A^{2}}$, $\frac{T_{\mu\nu}^{eff}}{B^{2}}$, $\frac{T_{\mu\nu}^{eff}}{AB}$, and $\frac{T_{\mu\nu}^{eff}}{C^{2}}$ be denoted by $k_{1}$, $k_{2}$, $k_{3}$ and $k_{4}$ respectively for notational simplicity, where
\begin{align}
k_{1} & =\frac{1}{f_{R}}\left[\rho+\epsilon+\epsilon f_{T}-\frac{\left(f-Rf_{R}\right)}{2}-\frac{D_{00}}{A^{2}}+\frac{3\varphi'^{2}}{A^{2}B^{2}}\right]\\
k_{2} & =\frac{1}{f_{R}}\left[\left(1+f_{T}\right)\left(p_{r}+\epsilon+4\eta\sigma\right)+\rho f_{T}\right.\nonumber\\
& \left.+\frac{\left(f-Rf_{R}\right)}{2}-\frac{D_{11}}{B^{2}}-\frac{3\varphi'^{2}}{A^{2}B^{2}}\right]\\
k_{3} & =\frac{1}{f_{R}}\left[-\left(1+f_{T}\right)\left(q+\epsilon\right)-\frac{D_{01}}{AB}\right]\\
k_{4} & =\frac{1}{f_{R}}\left[\left(1+f_{T}\right)\left(p_{\perp}-2\eta\sigma\right)+\rho f_{T}\right.\nonumber\\
& \left.+\frac{\left(f-Rf_{R}\right)}{2}-\frac{D_{22}}{C^{2}}-\frac{\varphi'^{2}}{A^{2}B^{2}}\right]
\end{align}
The eigenvalues $\mu_{0}$, $\mu_{1}$, $\mu_{2}$ and $\mu_{3}$ are obtained from \eqref{eigenvalue} as follows :

\begin{align}
\mu_{0} & =\frac{k_{2}-k_{1}-\Delta}{2}\\
\mu_{1} & =\frac{k_{2}-k_{1}+\Delta}{2}\\
\mu_{2} & =\mu_{3}=k_{4}
\end{align}
where,
\begin{equation}
\Delta=\sqrt{\left(k_{1}+k_{2}+2k_{3}\right)\left(k_{1}+k_{2}-2k_{3}\right)}
\end{equation}
From the condition that $\Delta^{2}$ has to be positive definite
for a real value of $\Delta$, we have the condition,
\begin{equation}\label{NEC}
\vert k_{1}+k_{2}\vert-2\vert k_{3}\vert\ge0
\end{equation}
which is actually the null energy condition (NEC), stating that null vectors (which move along the path of photon) cannot carry negative energy density.
The weak energy condition (WEC) requires that
\begin{align}
-\mu_{0}\ge0;\\
-\mu_{0}+\mu_{i}\ge0
\end{align}
where, i =1,2,3. This gives us the following conditions :
\begin{align}
k_{1}-k_{2}+\Delta & \ge0
\end{align}
and,
\begin{equation}
2k_{4}-\left(k_{2}-k_{1}-\Delta\right)\ge0
\end{equation}
These conditions require that the matter density is always non-negative for timelike observers.
The dominant energy condition (DEC) requires that
\begin{equation}
\mu_{0}\le\mu_{i}\le-\mu_{0}
\end{equation}
which gives us
\begin{equation}
k_{1}-k_{2}\ge0
\end{equation}
and,
\begin{equation}
k_{1}-k_{2}+\Delta-2k_{4}\ge0
\end{equation}
These conditions require that the flow of energy and momentum cannot exceed the velocity of light.
The strong energy condition (SEC) requires that
\begin{align}
-\mu_{0}+\underset{i}{\Sigma}\mu_{i} & \ge0\\
-\mu_{0}+\mu_{i} & \ge0
\end{align}
which gives us
\begin{equation}
\Delta+2k_{4}\ge0
\end{equation}
The SEC requires that gravity is always attractive, and the sum of energy and effective pressures in every direction is non-negative. These equations can be written as follows :

NEC :
\begin{align}\label{NEC1}
\vert\left(1+f_{T}\right)\left(\rho+p_{r}+4\eta\sigma+2\epsilon\right)-\frac{D_{00}}{A^{2}}-\frac{D_{11}}{B^2}\vert\nonumber\\
-2\vert\left(1+f_{T}\right) \left(q+\epsilon\right)+\frac{D_{01}}{AB}\vert\ge0
\end{align}

WEC :
\begin{align}
\label{WEC1}\frac{1}{f_{R}}\left[\rho\left(1-f_{T}\right)-\left(1+f_{T}\right)\left(p_{r}+4\eta\sigma\right)\right. &\nonumber\\
\left.-\left(f-Rf_{R}\right)-\frac{D_{00}}{A^{2}}+\frac{D_{11}}{B^{2}}+\frac{6Q^{2}}{C^{4}}\right]+\Delta &\ge0 \\
\label{WEC2}\frac{1}{f_{R}}\left[\left(1+f_{T}\right)\left(\rho-p_{r}+2p_{\perp}-8\eta\sigma\right)\right.\nonumber\\
\left.-\frac{D_{00}}{A^{2}}+\frac{D_{11}}{B^{2}}-\frac{2D_{22}}{C^{2}}+\frac{4Q^{2}}{C^{4}}\right]+\Delta\ge0
\end{align}

DEC :
\begin{align}
\label{DEC1}& \frac{1}{f_{R}}\left[\rho\left(1-f_{T}\right)-\left(1+f_{T}\right)\left(p_{r}+4\eta\sigma\right)-\left(f-Rf_{R}\right)\right. \nonumber\\
& \left.-\frac{D_{00}}{A^{2}}+\frac{D_{11}}{B^{2}}+\frac{6Q^{2}}{C^{4}}\right] \ge0 \\
\label{DEC2}& \frac{1}{f_{R}}\left[\rho-3\rho f_{T}-\left(1+f_{T}\right)\left(p_{r}+2p_{\perp}\right)-2\left(f-Rf_{R}\right)\right.\nonumber\\
& \left.-\frac{D_{00}}{A^{2}}+\frac{D_{11}}{B^{2}}+\frac{2D_{22}}{C^{2}}+\frac{8Q^{2}}{C^{4}}\right]+\Delta\ge0
\end{align}

SEC :
\begin{align}
\label{SEC1}& \frac{2}{f_{R}}\left[\left(1+f_{T}\right)\left(p_{\perp}-2\eta\sigma\right)+\rho f_{T}+\frac{\left(f-Rf_{R}\right)}{2}\right.\nonumber\\
& \left.-\frac{D_{22}}{C^{2}}-\frac{\varphi'^{2}}{A^{2}B^{2}}\right]+\Delta\ge0
\end{align}
For the inclusion of bulk viscosity, the radial and tangential pressure components will get reduced by the effect of the bulk viscosity term. The possibility of an overall negative pressure owing to the presence of a bulk viscosity term, may lead to situations where the entire left hand side of the inequality given by the SEC becomes negative. In that case, the SEC is violated, and hence, gravity will become repulsive. This violation of SEC in case of $f(R,T)$ gravity may provide a possible explanation of the currently observed accelerated expanding phase of the universe.

Choosing a linear form of the $f(R,T)$ function, where $f(R,T)=R+\lambda T$, equation \eqref{NEC1}, which represents the NEC, can be plotted against the energy density $\rho$ and the radial pressure $p_{r}$ for various values of $\eta$, $q$, $\sigma$ and $\epsilon$.
In terms of the structure scalars, the terms $k_{1}$, $k_{2}$, and
$k_{4}$ can be expressed as follows :

\begin{align}
k_{1} & =X_{T}\\
k_{2} & =\frac{1}{3}\left(2Y_{T}-2X_{TF}-2Y_{TF}-X_{T}\right)\\
k_{4} & =\frac{1}{3}\left(X_{TF}+Y_{TF}+2Y_{T}-X_{T}\right)
\end{align}
The term $k_{3}$ cannot be expressed in terms of any structure scalar.

Expressed in terms of the structure scalars, the energy conditions
now take the following form :

WEC :
\begin{align}
\frac{2}{3}\left(2X_{T}-Y_{T}+X_{TF}+Y_{TF}\right)+\Delta & \ge0\\
\frac{2}{3}\left(X_{T}+Y_{T}+2X_{TF}+2Y_{TF}\right)+\Delta & \ge0
\end{align}

DEC :
\begin{align}
2X_{T}-Y_{T}+X_{TF}+Y_{TF} & \ge0\\
2\left(X_{T}-Y_{T}\right)+\Delta & \ge0
\end{align}

SEC :
\begin{equation}
\frac{2}{3}\left(2Y_{T}-X_{T}+X_{TF}+Y_{TF}\right)+\Delta\ge0
\end{equation}

\section{Results and Conclusions}

In this paper, we have investigated the role of the structure scalars on the various matter variables for the collapse of a charged spherically symmetric shearing dissipative fluid in $f(R,T)$ gravity. The most general spherically symmetric spacetime has been considered for the interior of the collapsing matter. The structure scalars have been derived in terms of the spacetime geometry, and then, with the help of the $f(R,T)$ field equations, these have been related to the physical parameters of the matter, such as, energy density inhomogeneity, pressure anisotropy, shear viscosity, and charge. Initially, a general treatment has been provided without specifying the functional form of the $f(R,T)$ function, and later, the relation between the structure scalars and the matter variables have been provided for a linear form of the $f(R,T)$ function. We have also provided these relations in the special case of constant $R$ and $T$, for a relativistic dust ball. Choosing the exterior spacetime as the generalized Vaidya metric of outgoing radiation, the $f(R,T)$ junction conditions have been used to examine the relation between the radial pressure, heat flux, shear viscosity at the fluid boundary in presence of charge and modified gravity effects. Further, the relation between the matter Lagrangians of the exterior and the interior regions, along with their derivatives have also been found, in presence of charge and modified gravity effects. The energy conditions have been presented and the possibility of violation of the SEC in the context of $f(R,T)$ gravity has been discussed.

The following can be seen from the relation between the structure scalars and the matter variables given in equations \eqref{xt1}, \eqref{xtf1}, \eqref{yt1} and \eqref{ytf1} :
\begin{itemize}
\item In absence of dissipation, the energy density inhomogeneity is influenced by $X_{TF}$ and the mass-function $m$ together. As it can be seen from equation \eqref{xtfmodified} in the presence of electromagnetic field, the charge plays a role in influencing the energy density inhomogeneity and $X_{TF}$.
\item The heat dissipation and the energy density inhomogeneity is also controlled by $Z$ in addition to $X_{TF}$, as is evident from equation \eqref{xtfmodifiedwithz}.
\item The evolution of the expansion scalar is controlled by $Y_{T}$, as can be seen from equation \eqref{ytmodified}. The contribution of the charge is manifested in the structure scalar $Y_{T}$ as is evident from equation \eqref{yt2}.
\item The evolution of the shear scalar is controlled by $Y_{TF}$, as can be seen from equation \eqref{ytfmodified}. The contribution of the charge is present in the structure scalar $Y_{TF}$, as can be seen from equation \eqref{ytf2}.
\item The effective homogeneous energy density is influenced by $X_{T}$, as can be seen from equation \eqref{xtrhoeff}.
\item The pressure anisotropy is also influenced by both $Y_{T}$ and $Y_{TF}$.
\item The electric charge $Q$ appears in the expressions for all four of the structure scalars, causing an increment in $X_{T}$ and $Y_{TF}$, and a decrement in $X_{TF}$ and $Y_{T}$.
\item The presence of electric charge also causes an increase in the mass-energy content of the collapsing matter, as is evident from equation \eqref{massfn1}.
\item{The structure scalar $Y_{TF}$ can also be considered as the complexity factor, as it vanishes in case of isotropic pressure and homogeneous energy density in absence of modified gravity. The condition for vanishing complexity has also been obtained in equation \eqref{complexitycondition}, for which $Y_{TF}$ becomes zero.}
\end{itemize}

To the best of our knowledge, although the influence of electromagnetic field on the structure scalars had previously been discussed in the context of GR \cite{HerreraPRD2011}, $f(R)$ gravity \cite{BhattiPhysScr2021}, $f(R,T,Q)$ gravity \cite{YousafBhattiNaseer2020}, and similar other theories, but it was not done for $f(R,T)$ gravity. Unlike the case in GR \cite{HerreraPRD2011}, we have not absorbed the contribution from the charge into the effective matter variables (superscripted by $\mathscr{P}$), but kept it as a separate entity, in order to clearly identify the increment and decrement of the structure scalars due to the presence of the electromagnetic factor, which, as it can also be seen from equations \eqref{xt2}-\eqref{ytf2}, differ by their integer coefficients. This is a significant outcome of our study.

For the choice of an $f(R,T)$ function of the form $f(R)+\lambda T$, where $f(R)=R+\alpha R^{2}$, which is the Starobinsky model \cite{Astashenok1, Astashenok2, Mardan2025}, it can clearly be seen from the analysis of the junction conditions, that the higher order curvature terms in this case will play a significant role in constraining the matching conditions for the interior and exterior matter-Lagrangians for the collapse to be feasible, which in case of a linear functional model like $R+\lambda T$, and absence of charge and shear, will just result in the continuity of the matter Lagrangian across the boundary. Also for the case of a constant scalar curvature $R_{0}$, the higher order curvature terms would vanish. Black hole solutions in the presence of the cosmological constant can be obtained for constant scalar curvature and constant trace choices of the $f(R,T)$ function, in a way similar to the $f(R)$ theory case described in \cite{SharifAPSS2011}. Further, like the $f(R)$ junction conditions discussed in \cite{CasadoTurrion2022}, we have discussed the consequences of the $f(R,T)$ junction conditions which have the additional constraints of matching the Ricci scalar and its normal derivative, and the Trace of the energy-momentum tensor and its normal derivative across the boundary of the collapsing sphere for a smooth matching. These extra conditions have led us to the constraints on the interior and exterior matter-Lagrangians given by equations \eqref{lagrangiancondition}, \eqref{tderivative} and \eqref{rderivative}. Further the choice of our exterior spacetime as the generalized Vaidya spacetime for outgoing radiation has led to the further constraint on the choice of our $f(R,T)$ function which is given by equation \eqref{fRTconstraint}.

We have plans to extend this piece of investigation to other configurations of collapsing matter in modified gravity for different exterior spacetimes, and also for different geometry.

\section*{Acknowledgements}
The authors are thankful to the anonymous reviewer for the valuable comments and suggestions. UG is thankful to St. Xavier's College, Kolkata, India, for their Research Cell facilities. SG thanks IUCAA, India for the visiting associateship. A portion of this work was done in IUCAA, India under the associateship program. SG thanks Prof. Sunil D. Maharaj for helpful suggestions.

\end{document}